\documentclass{nature}
\usepackage[T1]{fontenc}
\usepackage[utf8]{inputenc}
\usepackage{graphicx}
\usepackage{float}
\usepackage{amssymb}
\usepackage{amsmath}
\usepackage{dcolumn}
\usepackage{color}
\usepackage{bm}
\usepackage[normalem]{ulem}
\usepackage[section]{placeins}
\usepackage[labelfont=bf]{caption}
\usepackage{enumitem}   
\usepackage{chemformula}
\usepackage{pdfpages}

\newcommand{\beq}{\begin{equation}}
\newcommand{\eeq}{\end{equation}}
\newcommand{\beqn}{\begin{eqnarray}}
\newcommand{\eeqn}{\end{eqnarray}}

\begin{document}
\title{Quarter- and half-filled quantum Hall states and their topological orders revealed by daughter states in bilayer graphene}
\author{Ravi Kumar$^{1,}$\footnote{These authors contributed equally: Ravi Kumar, Andr\'e Haug}, Andr\'e Haug$^{1,*}$, Jehyun Kim$^1$, Misha Yutushui$^1$, Konstantin Khudiakov$^1$, Vishal Bhardwaj$^1$, Alexey Ilin$^1$, K. Watanabe$^{2}$, T. Taniguchi$^{3}$, David F. Mross$^1$, and Yuval Ronen$^{1}$\footnote{yuval.ronen@weizmann.ac.il}}

\maketitle

\begin{affiliations}
\item Department of Condensed Matter Physics, Weizmann Institute of Science, Rehovot 76100, Israel.
\item Research Center for Functional Materials, National Institute for Materials Science, 1-1 Namiki, Tsukuba 305-0044, Japan.
\item International Center for Materials Nanoarchitectonics,
National Institute for Materials Science, 1-1 Namiki, Tsukuba 305-0044, Japan.
\end{affiliations}

\begin{abstract}
Even-denominator fractional quantum Hall states are promising candidates for fault-tolerant quantum computing due to their underlying non-Abelian topological order. However, the topological order of these states remains hotly debated. Here, we report transport measurements on ultra-clean bilayer graphene heterostructures, where we observed four quarter-filled states and their corresponding Levin–Halperin daughter states, constraining their topological order. Moreover, we complete the sequence of half-filled plateaus by detecting states at $\nu=-\frac{3}{2}$ and $\nu=\frac{1}{2}$ whose daughters suggest an alternating sequence of non-Abelian orders. This pattern suggests a universal origin supporting their use in identifying topological order at even-denominator fillings, though further confirmation is needed via direct measurements. The observed quarter- and half-filled states appear in $N=0$ and $N=1$ Landau levels, respectively, and thus highlight a competition between interactions favoring paired states of either four- or two-flux composite fermions. Additionally, we observe several ‘next-generation’ quantum Hall states that require strong interactions between composite fermions. 
\end{abstract}

\noindent\textbf{Introduction.}
The search for non-Abelian anyons has inspired groundbreaking research into the nature of the even-denominator fractional quantum Hall (FQH) states.\cite{Read_paired_2000} However, decades after the initial observation of a plateau at filling factor $\nu=\frac{5}{2}$ in GaAs,\cite{Willett_observation_1987} its nature and origin remain under intense debate.\cite{ma_fractional_2022} This enduring mystery stems from a rich competition between multiple exotic quantum states, which includes several sought-after non-Abelian states\cite{Moore_nonabelions_1991, Greiter_half_filled_1991, Lee_particle_hole_2007, Levin_particle_hole_2007, Son_is_2015} contending with more conventional Abelian states\cite{Halperin_QH_1983, Read_paired_2000} and metallic composite-fermion liquids.\cite{halperin_theory_1993} Recent observations of plateaus at half filling in different semiconductor devices\cite{suen1992observation,eisenstein1992new,suen1994origin,shabani2013phase, Falson_Zno_2015, singh_Topological_2024} and van der Waals materials\cite{Ki_Observation_2014, Li_Even_2017, Zibrov_Tunable_bilayer_graphene_2017, kim2019even,shi2020odd, Huang_Valley_2022, chen2024tunable} hint at diverse non-Abelian phases whose origins are yet to be unveiled.

Composite fermions (CFs), electrons bound to a pair of flux quanta and carrying a charge of $q_\text{CF} =(1-2 \nu) e$, are pivotal for explaining half-filled states.\cite{Jain_composite-fermion_1989} At half filling, CFs are charge neutral and impervious to magnetic fields. Consequently, they can form a metallic Fermi surface,\cite{halperin_theory_1993} which is commonly observed in the half-filled zeroth Landau level (LL), denoted by $N=0$. At the same filling, incompressible states develop when the metallic state transitions into a CF superconductor. The attraction necessary for CF pairing arises in the first excited LL ($N=1$), whose single-electron in-plane wave functions differ from those at $N=0$.\cite{Morf_transition_1998} Alternatively, pairing can be promoted in wide quantum wells where wave functions extend out-of-plane.\cite{Zhang_finite_thick_1986, Papic_wide_well_2009} However, the choice of the pairing channel---and thus, whether the state exhibits Abelian or non-Abelian characteristics---is influenced by subtle effects such as Landau-level mixing (LLM). Even more elusive are states formed at quarter fillings, previously reported via magneto-resistance measurements on GaAs\cite{Luhman_Observation_2008, Shabani_Evidence_2009, Shabani_Correlated_2009, Wang_Even_2022, Wang_Next_2023} and in a capacitance measurement on monolayer graphene.\cite{Zibrov_Even_Denominator_2018} These states comprise charge-neutral pairs of four-flux composite fermions, \textsuperscript{4}CFs, which condense into a superfluid, similar to the case at half filling.\cite{Zhao_CF_pairing_2023, Sharma_quarter_2024}

Attractive interactions at half filling are detrimental to FQH states of the Jain sequence at  $\nu=\frac{n}{2n\pm1}$.\cite{singh_Topological_2024} At these fillings, the CF charge $q_\text{CF}$ is non-zero. Consequently, these CFs are sensitive to magnetic fields, and the CF metal exhibits Shubnikov--de Haas (SdH) oscillations, which manifest as incompressible Jain states.\cite{Jain_Incompressible_1989} When a pairing gap forms on the composite Fermi surface, the SdH oscillations are suppressed, and even lower-order Jain states vanish around half-filled plateaus.\cite{Willett_observation_1987, singh_Topological_2024} A compromise between pairing and adjusting to the magnetic field takes the form of `daughter' states, first proposed by Levin and Halperin.\cite{Levin_Collective_2009} In these states, paired CFs enter bosonic quantum Hall states instead of condensing, resulting in FQH states at distinct filling factors determined by the pairing channel.\cite{Yutushui_daughters_2024, Zheltonozhskii_daughters_2024} This unique parent--daughter correspondence may shed light on the topological order at half filling and was used to identify states in bilayer graphene (BLG)\cite{Huang_Valley_2022, Assouline_energy_gap_2024} and GaAs.\cite{Kumar.2010.Nonconventional, singh_Topological_2024} However, this approach warrants confirmation through direct observations at half filling, such as via thermal conductance measurements.

BLG has established itself as a versatile platform for studying the superconductivity of both electrons\cite{zhou_isospin_2022} and CFs in even-denominator states.\cite{Ki_Observation_2014, Li_Even_2017, Zibrov_Tunable_bilayer_graphene_2017, Huang_Valley_2022} In contrast to GaAs, where $N=0$ and $N=1$ are separated by the cyclotron energy, these levels are degenerate in BLG. Here, the single-particle electron wavefunction is confined to a single layer when $N=0$, whereas, for $N=1$ it extends across both layers (Fig.~1A, left), reducing Coulomb repulsion and facilitating electron pairing.\cite{McCann_BLG_2006} This degeneracy, combined with additional degeneracies in spin ($\sigma=\uparrow,\downarrow$) and valley ($K,K^\prime$) quantum numbers, gives rise to a total of 8 degenerate ground states. In high-mobility heterostructures measured at low temperatures and high magnetic fields, these degeneracies are lifted, revealing a phase space spanned by the displacement field and filling factor, with 16 crossings of integer LLs.  (Fig.~1A, right).\cite{Hunt_Direct_measurement_2017, Li_Even_2017,xiang2023intra,Kousa_orbital_2024} By exploring this phase space, plateaus at multiple distinct half-integer fillings have been identified as either Pfaffian or anti-Pfaffian according to their daughter states\cite{Huang_Valley_2022,Assouline_energy_gap_2024}. However, the periodic pattern of non-Abelian states remained hidden due to missing plateaus at $\nu=-\frac{3}{2}$ and $\nu=\frac{1}{2}$. Quantum Hall states at quarter-integer fillings and other `next-generation'\cite{Smet2003,chang_generations_2004, Wang_Next_2023} even denominators had not been seen in BLG, to the best of our knowledge.

In this work, we report the observation of incompressible FQH states at quarter- and half-filled LLs in BLG, which we attribute to a combination of high sample quality, highly transparent contacts, and an ultra-low noise measurement setup (see Methods section). First, our measurement of additional plateaus at $\nu=-\frac{3}{2}$ and $\nu=\frac{1}{2}$ reaffirms an orbital index of $N=1$ as a prerequisite to forming incompressible states at half fillings. By observing their daughters, we identify both states as anti-Pfaffian. These plateaus complete the sequence of previously known states and reveal a systematic alternation between anti-Pfaffian and Pfaffian in odd and even LLs, respectively. Moreover, our work directly links the half-filled plateaus and their daughters to Landau level crossings, which introduced a perspective that was not addressed in previous works. Second, we observe the emergence of incompressible states at four distinct quarter-integer fillings, with plateaus appearing exclusively within the $N=0$ level, consistent with previous studies.\cite{Luhman_Observation_2008, Shabani_Evidence_2009, Shabani_Correlated_2009, Wang_Even_2022, Wang_Next_2023}. Moreover, by resolving their daughter states, we were able to constrain their topological order. The formation of quarter-filled states at $N=0$, in contrast to half-filled states at $N=1$, indicates a competition between the interactions favoring either paired states of \textsuperscript{4}CFs or of \textsuperscript{2}CFs, respectively. Finally, we observed odd and even `next-generation' FQH states at partial fillings of \textsuperscript{2}CFs.\\
\noindent\textbf{Results.}
Our devices (device 1 and device 2) consist of BLG encapsulated by hexagonal boron nitride and graphite layers acting as gates;\cite{pizzocchero2016hot} see Methods section. Subsequently, we fabricated the heterostructures into a Hall bar, as shown in supplementary note~1. The electrical transport properties of device 1 were measured in a dilution refrigerator with a base temperature $T=16~\mathrm{mK}$ and a magnetic field $B$ up to $18~\mathrm{T}$. Device 2 was measured in another dilution refrigerator with a base temperature $T=10~\mathrm{mK}$ and a magnetic field up to $12~\mathrm{T}$. The data shown in the main manuscript are from device 1 and are consistent with those of device 2 (see supplementary note~12). Using a standard low-frequency lock-in technique, we measured the longitudinal ($R_{xx}$) and transverse ($R_{xy}$) resistances; see supplementary note~1. Subsequently, fully developed FQH states were identified according to their $R_{xy}$ values and accompanying minima in $R_{xx}$. For less developed plateaus, the filling factors were determined by the position of their $R_{xx}$ minima with respect to fully developed plateaus, e.g., $\frac{2}{5}$, $\frac{1}{2}$, and $\frac{3}{5}$, etc. The dual-gate geometry permits independent tuning of filling factor $\nu$ and displacement field $D$, thus granting control over both the orbital ($N=0,1$) and the valley isospin ($K$, $K^\prime$).\cite{Hunt_Direct_measurement_2017, Li_Even_2017} 

Fig.~1B shows $R_{xx}$ and $R_{xy}$  as a function of $\nu$ at $D=-80~\mathrm{mV/nm}$ and $B=18~\mathrm{T}$, where the integer LL has an $N=0$ orbital. We observe a compressible \textsuperscript{2}CF metal at $\nu=\frac{1}{2}$, flanked by an abundance of SdH oscillations, testifying to the low disorder of the device. As we increase the displacement field to $D=-160~\mathrm{mV/nm}$, the LL transitions to $N=1$. Here, we observe an incompressible state at $\nu=\frac{1}{2}$ and a strong suppression of the Jain states; see Fig.~1C. The diminished $R_{xx}$ along with a quantized plateau in $R_{xy}$ indicates pairing of \textsuperscript{2}CFs at $\nu=\frac{1}{2}$. Surprisingly, near $\nu=\frac{3}{4}$, the roles of $N=0$ and $N=1$ are reversed: For $N=0$, we observe a strong suppression of $R_{xx}$ along with a plateau in $R_{xy}$ at $\nu=\frac{3}{4}$, indicating the pairing of \textsuperscript{4}CFs.
By contrast, at $N=1$, this filling hosts a compressible \textsuperscript{4}CF metal flanked by SdH oscillations. Fig.~1D shows $R_{xx}$ over a continuous range of $\nu$ and $D$ which confirms the robustness of both half- and quarter-filled states. Moreover, it shows that these states form irrespective of the valley isospin (see supplementary note~4 Fig.~9). The quarter- and half-filled states at $\nu=\frac{3}{4}$ and $\nu=\frac{1}{2}$ were also observed in device 2, see supplementary note~12. Consistent with previous studies,\cite{Li_Even_2017, Huang_Valley_2022} which reported a slope of $2\times 10^{-4}$$ \frac{1}{\mathrm{mV/nm}}$ in the minima of $R_{xx}$ within the $\nu$--$D$ phase space, our observation revealed a slope of approximately $3\times 10^{-4}\frac{1}{\mathrm{mV/nm}}$. These slopes are most likely due to changes in quantum capacitance, which changes with displacement field (see supplementary note~4 for more discussion on this point).

\begin{figure}[hbt!]
    \includegraphics[width=\textwidth]{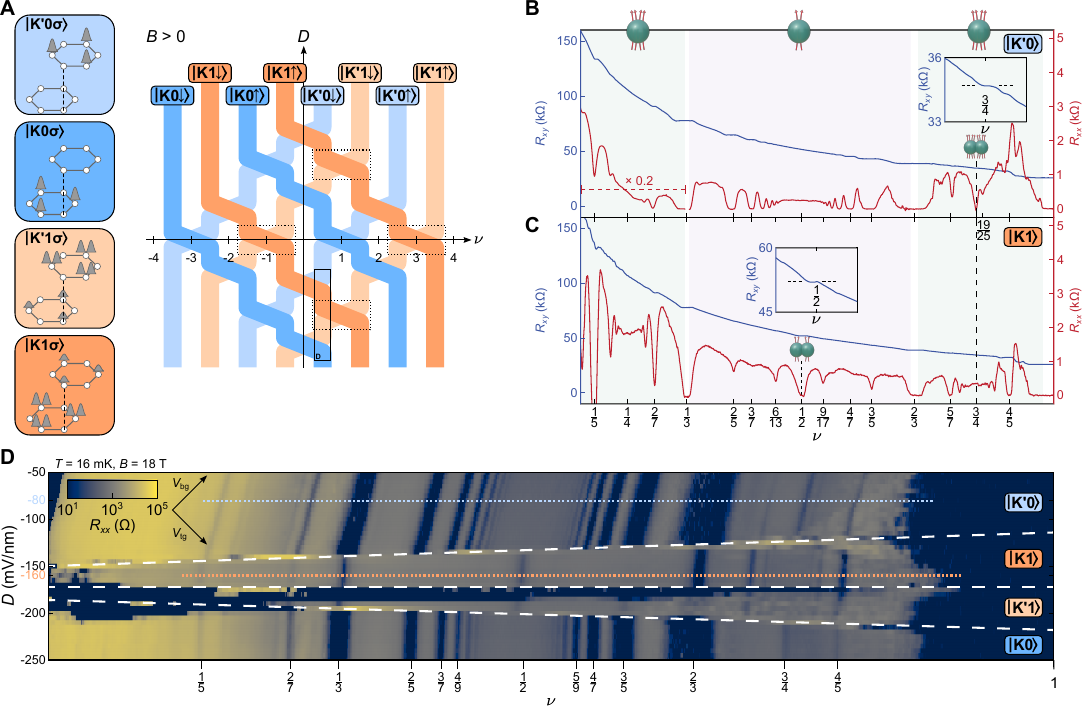}
    \caption{\textbf{Interplay between quarter- and half-filled FQH states in BLG.} (\textbf{A}) BLG unit cell's real space single-electron wave functions for the zero-energy LL, eightfold degenerate in valley isospin, orbital, and spin (left) and the evolution of LLs as their degeneracy is lifted (right). Dotted boxes refer to crossings of $N=1$ LLs. The solid black line box refers to level crossings observed in panel D. (\textbf{B},\textbf{C}) Longitudinal $R_{xx}$ (red) and transverse $R_{xy}$ (blue) resistances as a function of $\nu$ at $T=16~\mathrm{mK}$, $B=18~\mathrm{T}$ for $D=-80~\mathrm{mV/nm}$ ($N=0$) (B) and $D=-160~\mathrm{mV/nm}$ ($N=1$) (C). Panels in the top right corners indicate the valley isospin and orbital. The shaded purple (center) and green (sides) regions correspond to FQH states composed of \textsuperscript{2}CFs and \textsuperscript{4}CFs. The insets show zoom-ins to $R_{xy}$ plateaus at $\nu=\frac{1}{2}$ (C) and $\nu=\frac{3}{4}$ (B). We attribute these plateaus to pairing of \textsuperscript{2}CFs and \textsuperscript{4}CFs, respectively. In (B), $R_{xx}$ data below $\nu=\frac{1}{3}$ were divided by $5$ to accommodate large $R_{xx}$ values. (\textbf{D}) $R_{xx}$ as a function of $\nu$ and $D$, controlled electrostatically by altering the back and top gate voltages, $V_\mathrm{bg}$ and $V_\mathrm{tg}$. The dotted horizontal lines at $D=-80~\mathrm{mV/nm}$ (light blue) and $D=-160~\mathrm{mV/nm}$ (orange) indicate the line cuts displayed in (B) and (C), respectively.}
    \label{fig1}
\end{figure}

We now broaden our study to all eight levels comprising the zero-energy LL. In Fig.~2A, we show $R_{xx}$ as a function of $\nu$ and $D$ for the hole-doped (left) and electron-doped (right) sides. Dark-blue areas mark $R_{xx} < 10~\Omega$ where the bulk becomes incompressible. On the hole (electron) side, the $R_{xx}$ measurement is overshadowed by artifacts for negative (positive) $D$ due to the contact fabrication scheme; see Methods section. We observed all previously reported half-filled states at $\nu = -\frac{5}{2}, -\frac{1}{2}, \frac{3}{2}, \frac{5}{2}, \frac{7}{2}$ along with the same daughter states\cite{Huang_Valley_2022}. All theoretically predicted daughter states occur at filling factors that coincide with the Jain sequence $\nu_\text{Jain}=\frac{n}{2n+1}$ for relatively large $|n|$. However, the gaps of the Jain sequence are expected to decrease as $\Delta\sim \frac{1}{2n+1}$ toward half filling, consistent with observations in $N=0$ LLs.\cite{jain_composite_2007, singh_Topological_2024} We identify anomalously strong states that violate this pattern as daughter states, by comparing the strength of the $R_{xx}$ minima. For example, in Fig.~1C, we observe Jain states only up to $n=3$ ($\nu = \frac{3}{7}$) on the particle-like side and down to $n=-4$ ($\nu=\frac{4}{7}$) on the hole-conjugate side. This observation strongly suggests that the prominent states at $n=6$ ($\nu = \frac{6}{13}$) and $n=-9$ ($\nu=\frac{9}{17}$) have a fundamentally different character, i.e., they are daughter states.

The half-filled state at $\nu=-\frac{3}{2}$ emerges in a single pocket around $D=0$, accompanied by daughter states at $\nu=-\frac{20}{13}, -\frac{25}{17}$ (see supplementary note~10 Fig.~21). The one at $\nu=\frac{1}{2}$ occurs in two pockets centered around $D = \pm 170~\mathrm{mV/nm}$, with daughters at $\nu=\frac{6}{13},\frac{9}{17}$. All half-filled states develop only at $N=1$, where the Jain states are strongly suppressed, and quarter-filled states are absent.
Upon tuning the orbital index from $N=1$ to $N=0$, we observed quarter-filled states at $\nu = \frac{3}{4} +(-4,-2,0,+2)$ and a suppression of the \textsuperscript{4}CF Jain sequences around these fillings. This observation is consistent with previous studies in GaAs.\cite{Shabani_Evidence_2009,Luhman_Observation_2008,Shabani_Correlated_2009,Wang_Next_2023,Wang_Fractional_2023} Here, quarter-filled states in hole-doped systems were attributed to LLM, estimated to be $\frac{E_\mathrm{C}}{\Delta}\approx 3-8$, where $E_C$ is the Coulomb energy and $\Delta$ is the cyclotron energy. The estimated LLM in BLG is similar, i.e., $\frac{E_\mathrm{C}}{\Delta_{01}}\approx 5-10$, with $\Delta_{01}$ the gap between the $N=0,1$ levels of the same spin and valley,\cite{Zibrov_Tunable_bilayer_graphene_2017} and could therefore promote \textsuperscript{4}CF pairing. More importantly, we observe their daughter states at $\nu=\frac{9}{25}+(-4,-2,0)$; see Fig.~2C--F, which we use to constrain their topological order. The systematic appearance of these states in every second LL in BLG along with identical daughters suggests a common origin. \\
In Fig.~2B, we summarize the observed quarter- and half-filled states along with their daughter states, highlighting their dependence on the orbital and valley isospin indices. The half- and quarter-filled states that were not observed in previous works\cite{Li_Even_2017,Zibrov_Tunable_bilayer_graphene_2017,Huang_Valley_2022}, are shown by bold, solid black lines; see supplementary Figs.~11 and 12 for an enlarged view of hole and electron side along with their schematics. We already labeled the even-denominator states according to their identification via Levin--Halperin daughter states and in-plane $B$ measurements, which we discuss below.

\begin{figure}[H]
    \centering
    \includegraphics[width=\textwidth]{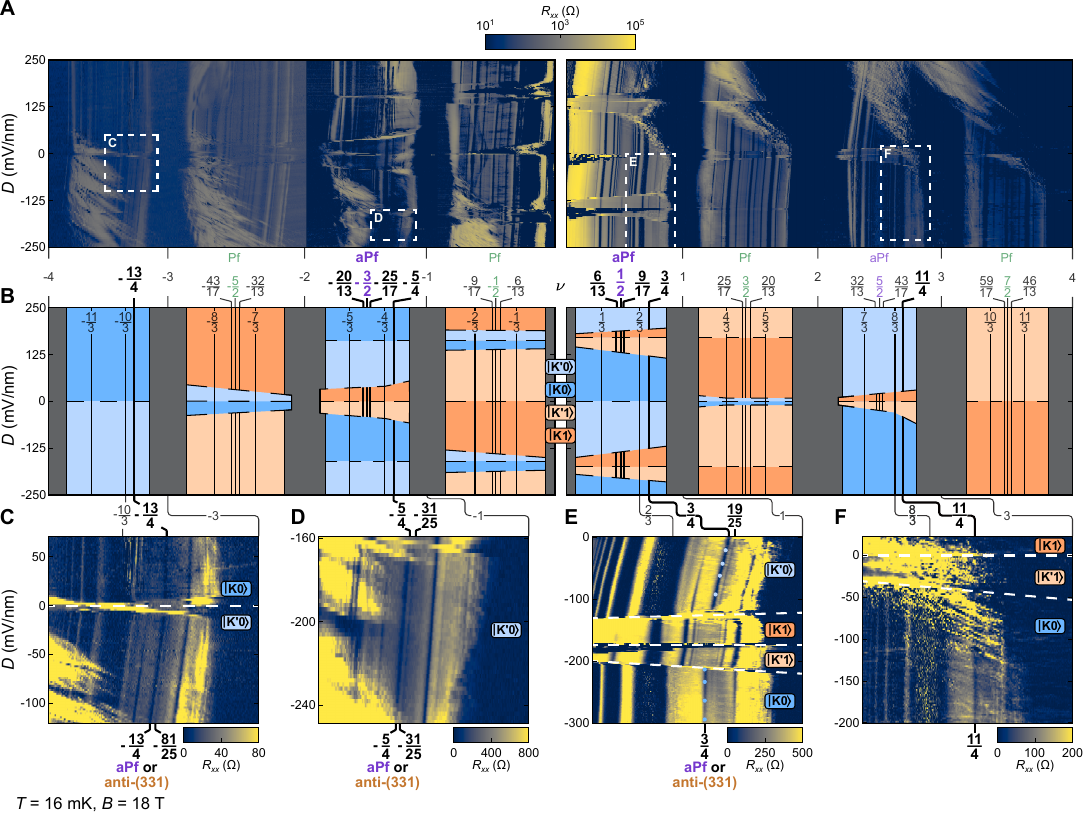}
    \caption{\textbf{Even-denominator FQH states in the zero-energy LL of BLG.} (\textbf{A}) $R_{xx}$ as a function of $\nu$ and $D$ for $-4\leq\nu<0$ (left panel) and $0<\nu\leq4$ (right panel). Areas highlighted by dashed boxes refer to high-resolution measurements in (C-F). See supplementary note~4 for enlarged view. (\textbf{B}) Schematic summary of  all even- and selected odd-denominator FQH states observed in (A), with higher-order Jain states omitted to increase visibility. Vertical lines mark observed FQH states; bold vertical lines mark FQH states at half- and quarter fillings that were not observed in previous works\cite{Li_Even_2017,Zibrov_Tunable_bilayer_graphene_2017,Huang_Valley_2022}. Dashed lines mark valley and orbital crossings along the $\nu$--$D$ phase space. Green and purple labels indicate Pfaffian (Pf) and anti-Pfaffian (aPf) states, respectively, according to their Levin--Halperin daughters. (\textbf{C--F})  $R_{xx}$ measurements near the four quarter-filled states. Blue dots in (E) mark the values of $D$ where we measured the gap $\Delta_{\frac{3}{4}}$ in Fig. 4E.}
    \label{fig2}
\end{figure}

We proceed with a detailed characterization of all half-filled states. To avoid ambiguity, we identify all states according to the topological phase of the \textit{electrons} for either sign of $\nu$. In Fig.~3A, we show the filling factors of daughters associated with different pairing channels of \textsuperscript{2}CFs.\cite{Yutushui_daughters_2024,Zheltonozhskii_daughters_2024} In Fig.~3B, we show $R_{xx}$ and $R_{xy}$ around $\nu=\frac{1}{2}$ at $D=-140~\mathrm{mV/nm}$, with strong dips in $R_{xx}$ flanking the plateau at half filling. The absence of higher-order Jain states beyond $\nu=\frac{2}{5},\frac{3}{5}$ suggests that the dips near $\nu=\frac{1}{2}$ are daughter states. We mark possible daughter-state fillings by vertical dashed lines with their colors referring to the topological order of the parent state from Fig.~3A, including Pfaffian and anti-Pfaffian favored by numerics\cite{ma_fractional_2022} and PH-Pfaffian suggested by thermal conductance measurements in GaAs.\cite{Banerjee_observation_2018, Dutta_Isolated_2022,Dutta_novel_2022}
The observed dips coincide with $\frac{6}{13}$ and $\frac{9}{17}$ (see supplementary note~9 for quantization of daughters), supporting anti-Pfaffian pairing for $\nu=\frac{1}{2}$. Similarly, based on the daughters, we identified the state at $\nu=-\frac{3}{2}$ to be anti-Pfaffian (supplementary note~10). Our analysis of the other half-filled states and the plateaus of their daughters, which agrees with previous reports,\cite{Huang_Valley_2022} is shown in supplementary notes~8 and 9. The identification of paired states based on their daughters is indirect and has not been independently confirmed by direct measurements, such as thermal conductance\cite{Banerjee_observation_2018, Dutta_Isolated_2022} or upstream noise at interfaces.\cite{Dutta_novel_2022} Still, the periodic pattern of topological orders indicated by daughters\cite{Yutushui_daughters_2024,Zheltonozhskii_daughters_2024} and its systematic association with crossing $N=1$ levels (see dotted boxes in Fig.~1A) provides strong support for its veracity.

Fig.~3C shows the dependence of the activation gap $\Delta_{\nu=\frac{1}{2}}$ on an in-plane magnetic field $B_\parallel$ at an out-of-plane field of $B_\perp = 15~\mathrm{T}$. The activation gap of the states examined in the manuscript is determined from the slope of the linear fit to the Arrhenius plot of $R_{xx}$ versus temperature. The error bars reflect the uncertainty in this linear fit (see supplementary note ~7 for more details). This method of determining the activation gap and its associated error bar is applied to all figures and tables in the manuscript. We observe negligible variations of $\Delta_{\frac{1}{2}}$ up to the largest accessible value of $B_\parallel = 10~\mathrm{T}$ which corresponds to a Zeeman energy of approximately $14~\mathrm{K}$, more than an order of magnitude larger than $\Delta_{\frac{1}{2}}$, similar to previous work.\cite{Li_Even_2017} We, therefore, conclude that the $\nu=\frac{1}{2}$ state is fully spin-polarized, as expected for the anti-Pfaffian. Note that, to fully validate the spin polarization, a direct approach such as NMR study is required.\cite{tiemann2012unraveling,stern2012nmr}  A detailed characterization of $\nu=-\frac{3}{2}$ also indicates a spin-polarized state; see supplementary note~10.\\
We summarize our identification of the half-filled states in the table of Fig.~3D (see supplementary notes~5 and 8 for gap measurements and daughters of other half-filled states). Strikingly, all observed anti-Pfaffians (aPf) occur in a pair with a Pfaffian (Pf) in the vicinity of two crossing $N=1$ levels. There is one such crossing for ($-\frac{3}{2}$,$-\frac{1}{2}$), two for ($\frac{1}{2}$,$\frac{3}{2}$), and one for ($\frac{5}{2}$,$\frac{7}{2}$), each exhibiting (aPf, Pf) on both sides of the crossing (see Fig.~1A and supplementary note~1~Fig.~2 for an illustration). This observation suggests that LLM within these pairs is most relevant for determining the topological order at half filling. Pfaffian or anti-Pfaffian are favored by LLM,\cite{Rezayi_breaking_2011,Pakrouski_phase_2015,Rezayi_Landau_2017} which is opposite for the two members of each pair. In the lower one, electrons can virtually occupy the empty level above; in the upper one, holes can virtually occupy the filled level below. By contrast, the spin and valley isospin appear to play subordinate roles. The largest gap occurred at $\nu=\frac{3}{2}$. Fig.~3E shows $R_{xx}$ around this filling as a function of $\nu$ and $B$. The incompressible state at $\nu=\frac{3}{2}$ persists down to $B=6~\mathrm{T}$, which places half-filled states in BLG within reach of commonly available magnetic-field setups.
\begin{figure}[H]
    \centering
    \includegraphics[width=\textwidth]{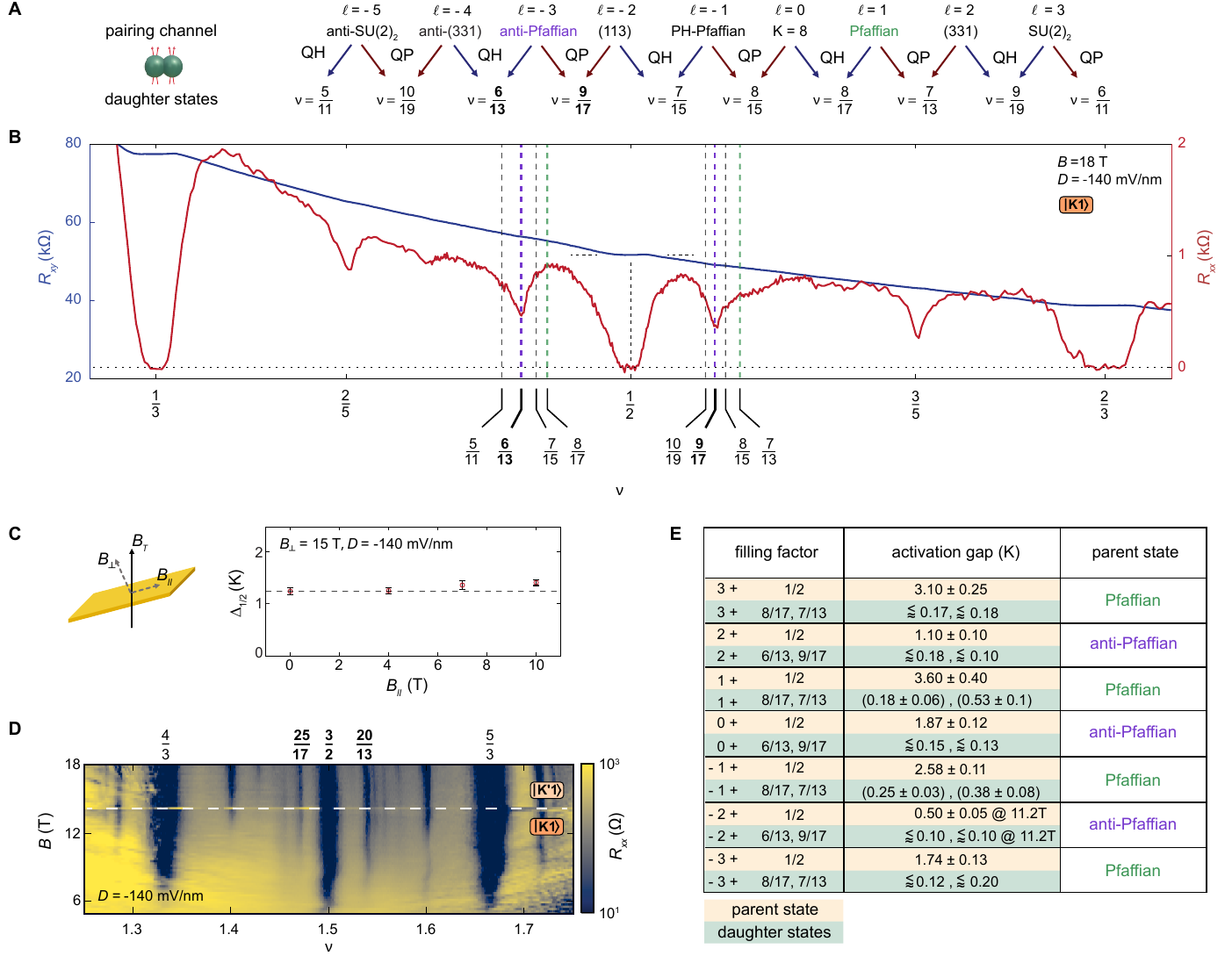}
    \caption{\textbf{Topological orders of half-filled states in BLG.} (\textbf{A}) Classification of half-filled states and their identification based on daughter-state fillings. The pairing channel $\ell$ corresponds to the number of Majorana edge channels (see supplementary note~13). (\textbf{B}) $R_{xx}$ and $R_{xy}$ around $\nu=\frac{1}{2}$, measured at $B=18~\mathrm{T}$ and $D=-140~\mathrm{mV/nm}$, along with possible daughter states marked by vertical dashed lines. (\textbf{C}) The thermal activation gap $\Delta_{\frac{1}{2}}$ measured as a function of $B_\parallel$ at $B_\perp=15~\mathrm{T}$. The dashed line represents the gap value at $B_\perp~=~15~\mathrm{T}$ and $B_\parallel=0~\mathrm{T}$. (\textbf{D}) Identification of all observed half-filled states based on their daughters, along with the measured thermal activation gaps. The gap of all the half-filled states was measured at $B=18~\mathrm{T}$, except for $\nu=-\frac{3}{2}$ which was measured at $B=11.2~\mathrm{T}$.(\textbf{E}) $R_{xx}$ around $\nu=\frac{3}{2}$ measured as a function of $B$ for $D=-140~\mathrm{mV/nm}$.}
    \label{fig3}
\end{figure}

Now we turn to a detailed analysis of the quarter-filled states. Similar to their cousins at half filling, quarter states can also exhibit characteristic daughters,\cite{Yutushui_daughters_2024} yet to be observed in other systems. In Fig.~4A, we show the filling factors of daughters associated with the main pairing channels. In Fig.~4B, we show $R_{xx}$ and $R_{xy}$ around $\nu=\frac{3}{4}$ with the possible daughter states indicated by dashed lines, with their colors referring to the topological order of the parent state from Fig.~4A. The line at $\nu=\frac{19}{25}$ shows remarkable agreement with our data, thereby suggesting the $\nu=\frac{3}{4}$ topological order to be the non-Abelian anti-Pfaffian or the Abelian anti-(331).
The absence of the daughter state to the left of $\nu=\frac{3}{4}$ is currently unclear and remains a subject of future investigation.

\begin{figure}[H]
    \centering
    \centerline{\includegraphics[width=\textwidth]{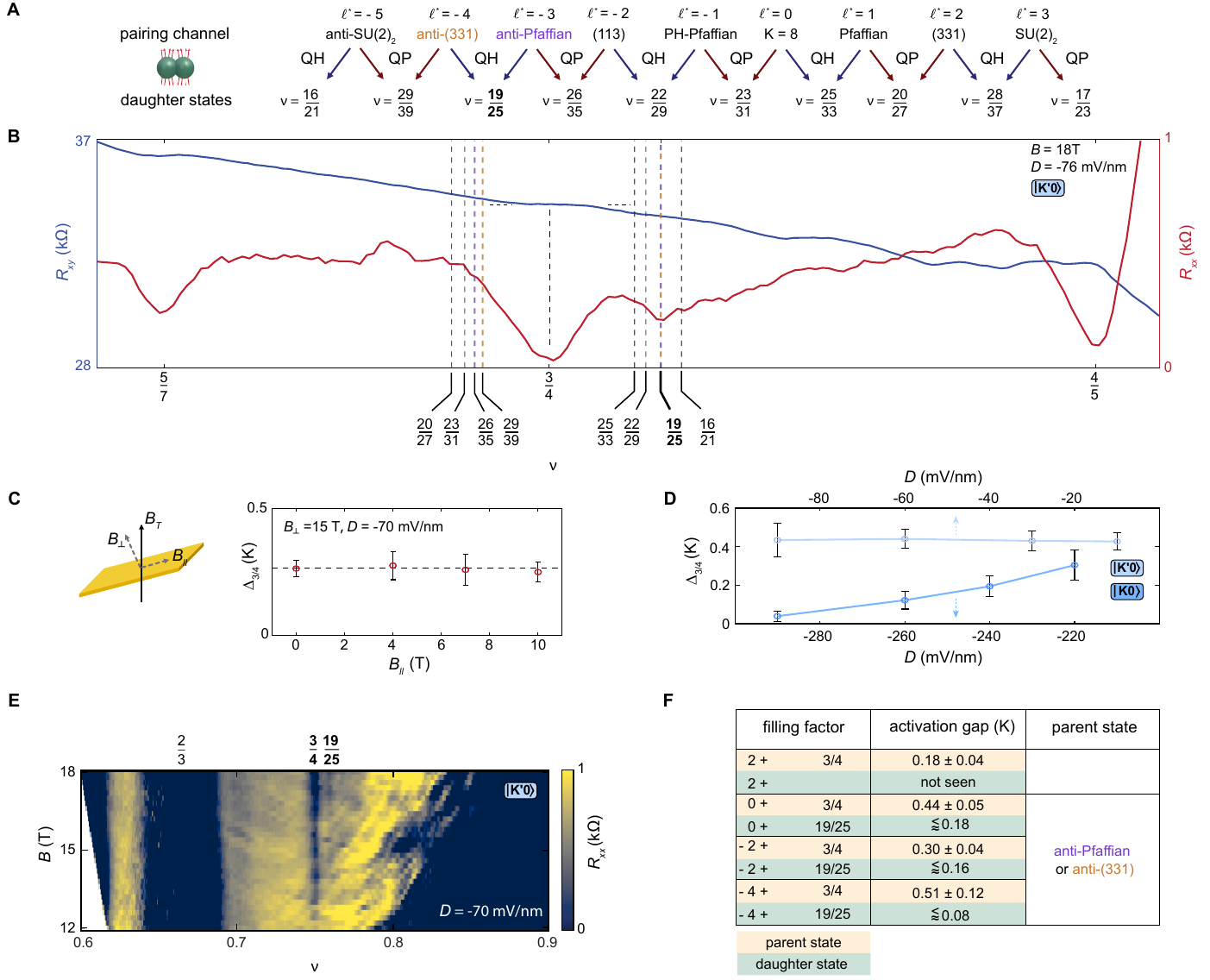}}
    \caption{\textbf{Quarter-filled states in BLG.} (\textbf{A}) Classification of quarter-filled states and their identification based on daughter-state fillings. The pairing channel $\ell$ corresponds to the number of Majorana edge channels (see supplementary note~13). (\textbf{B}) $R_{xx}$ and $R_{xy}$ around $\nu=\frac{3}{4}$ measured at $D=-76~\mathrm{mV/nm}$, along with possible daughter states marked by vertical dashed lines. (\textbf{C}) The thermal activation gap of $\nu=\frac{3}{4}$ measured as a function of  $B_\parallel$ at $B_\perp~=~15~\mathrm{T}$. The dashed line represents the gap value at $B_\perp~=~15~\mathrm{T}$ and $B_\parallel=0~\mathrm{T}$.
    (\textbf{D}) $\Delta_{\frac{3}{4}}$ measured as a function of $D$ in different valleys.
    (\textbf{E}) $R_{xx}$ around $\nu=\frac{3}{4}$ measured as a function of $B$ for $D=-70~\mathrm{mV/nm}$. 
    (\textbf{F}) Constraints on the ground states of all observed quarter-filled states based on their daughters, along with the measured thermal activation gaps.}
    \label{fig4}
\end{figure}
To determine the spin polarization of the $\nu=\frac{3}{4}$ state, we measured its gap while varying $B_\parallel$ at $B_\perp = 15~\mathrm{T}$; see Fig.~4C and supplementary note~7 for more details. The weak dependence of $\Delta_{\frac{3}{4}}$ on $B_\parallel$ up to a Zeeman energy of approximately $14~\mathrm{K}$ again indicates a fully spin-polarized state. In conjunction with the observed daughter state, this observation supports the non-Abelian anti-Pfaffian ground state, as the spin-singlet anti-(331) state should not be spin-polarized.\cite{Haldane_fqh_1983} However, a valley-singlet anti-(331) state remains possible.
To probe the valley polarization, we measured $\Delta_{\frac{3}{4}}$ as a function of $D$; see Fig.~4D and supplementary note~6 for more details. For the $K^\prime$ valley, we find that the gap remains constant as the absolute value of $D$ increases. By contrast, in the $K$ valley, the gap decreases monotonically. However, this behavior does not necessarily imply a valley singlet state. Indeed, a reduction of the gap with $D$ was also reported for $\nu=\frac{3}{2}$.\cite{Li_Even_2017} There, the gap eventually saturates to a non-zero value, and a single-component state is further supported by the daughters.\cite{Huang_Valley_2022} Consequently, our measurement can neither rule out an anti-Pfaffian nor a valley-singlet anti-(331) state. Fig.~4E shows $R_{xx}$ as a function of $\nu$ and $B$, with the incompressible state at $\nu=\frac{3}{4}$ persisting down to $B=12~\mathrm{T}$. We summarize our analysis of all observed quarter-filled states and their daughters in the table of Fig.~4F (see supplementary notes~5 and 8 for gap measurements and daughters of other quarter-filled states).

Finally, we have also observed `next-generation' FQH states previously unseen in BLG which are not captured by weakly interacting CFs.\cite{pan_fractional_2003, chang_generations_2004, Wojs_fractional_clustered_2004, pan_fractional_2015, Wang_Next_2023} Namely, we observed the odd-denominator states $\nu=\frac{4}{11}$ and $\nu=\frac{6}{17}$ and the even-denominator state $\nu = \frac{3}{8}$; see supplementary note~11 for details.

\noindent\textbf{Discussion.}
We uncovered a systematic and universal manner in which the topological orders of half-filled FQH states in BLG are determined. Near crossings of two $N=1$ levels, we consistently find (aPf, Pf) in the lower and upper levels, respectively, suggesting that LLM within these pairs is decisive. Additionally, we have observed four systematically occurring paired states of \textsuperscript{4}CFs and suggested their topological order based on daughter states to be of either anti-Pfaffian or anti-(331) order. The quarter-filled states arise exclusively in $N=0$ orbitals, oppositely to the half-filled states, which occur at $N=1$. Moreover, the $N=0$ orbitals also host `next-generation' odd- and even-denominator FQH states, previously only observed in GaAs\cite{pan_fractional_2003, chang_generations_2004,pan_fractional_2015, samkharadze_observation_2015} and in suspended monolayer graphene.\cite{kumar_unconventional_2018}

Our work highlights two arenas of competing CF interactions. First, their pairing competes against the formation of integer quantum Hall states of CFs, i.e., the Jain sequence. Second, interactions favoring the pairing of \textsuperscript{2}CFs disfavor pairing of \textsuperscript{4}CFs and vice versa. Both competitions can be observed and manipulated electrostatically in a single BLG heterostructure.
\pagebreak

\pagebreak
\section*{Methods}
The heterostructure was fabricated using the well-known dry-transfer stacking technique.\cite{pizzocchero2016hot} The flakes of hexagonal boron nitride (hBN), BLG, and few-layer graphite for the gates were exfoliated on a silicon substrate with a $285\text{-}\mathrm{nm}$ SiO\textsubscript{2} layer. A polydimethylsiloxane stamp covered with polycarbonate (PC) was used to pick up the top graphite flake and then, subsequently, hBN, BLG, hBN, and bottom graphite flakes from the substrate. The heterostructure, along with the PC film, was then landed on a clean substrate of the same kind. The substrate was cleaned in chloroform for two hours to remove residual PC. The heterostructure was then vacuum-annealed at $400^\circ\mathrm{C}$ for three hours and an AFM tip in contact mode was used to iron the heterostructure \cite{Purdie_Cleaning_2018} and determine the thicknesses of individual flakes. The desired geometry was achieved by dry-etching (O\textsubscript{2} for graphite gates, CHF\textsubscript{3}/O\textsubscript{2} for hBN).\cite{wang_one_2013} The BLG layer was then contacted using one-dimensional edge contacts.\cite{wang_one_2013} Special care was taken to improve the contact quality by optimizing the ratio of chromium ($2\text{--}3~\mathrm{nm}$), palladium ($13~\mathrm{nm}$), and gold ($70~\mathrm{nm}$) and depositing them at an angle of $\approx 15^{\circ}$ while continuously rotating the sample, leading to a low contact resistance of $\approx150~\Omega\mu\mathrm{m}$. After deposition, a lift-off procedure was performed in acetone and IPA. The final device's optical image is shown in supplementary Fig.~1A. Four-probe resistance measurements were performed using the standard lock-in technique. The measurement setup's noise level was monitored with a spectrum analyzer and minimized by removing many ground loops, adding low-pass filters both at room and low temperature and using Thermocoax wiring to shield high-frequency noise. An illustration of the measurement scheme is shown in supplementary Fig.~1B.

All measurements were performed in the $n$--$D$ phase space with the desired values converted to top and bottom gate voltages according to
\begin{align}
\begin{split}
    V_\mathrm{tg} &= \frac{d_\mathrm{hBN, top}}{\varepsilon_\mathrm{hBN}} \left(\frac{n e}{2\varepsilon_0} - D\right) + V_\mathrm{CNP, top},\\
    V_\mathrm{bg} &= \frac{d_\mathrm{hBN, bottom}}{\varepsilon_\mathrm{hBN}} \left(\frac{n e}{2\varepsilon_0} + D\right) + V_\mathrm{CNP, bottom},
    \end{split}
\end{align}
where $d_\mathrm{hBN,(top,bottom)}$ are the top and bottom hBN thicknesses, $\varepsilon_\mathrm{hBN} = 3.9$ is the dielectric constant of hBN, and $V_\mathrm{CNP,(top, bottom)}$ is the top and bottom gate voltage corresponding to the charge-neutrality point of the device.The larger slopes in our $R_{xx}$ data compared to previous publications\cite{zhou_isospin_2022,Li_Even_2017} can be traced back to an error of $0.7~\mathrm{nm}$ in one of the hBN thicknesses from the AFM measurement.

The device geometry reveals the origin of meandering artifacts in our $R_{xx}$ data. Due to the placement of the different gates, we obtain several interfaces at the edge: From the actual Cr/Pd/Au contact to a silicon-gated region on to another region that is gated by the graphite top gate and the silicon gate and then the actual device. At negative (positive) $D$ on the hole (electron) side, the silicon bottom gate is at a negative (positive) voltage, and the graphite top gate is at a positive (negative) voltage, thus depleting the contact region of charge carriers, resulting in large contact resistance and, eventually, the artifacts seen in our $R_{xx}$ data.

\section*{Data and materials availability:}
Data supporting the figures and tables of the main manuscript are available at https://doi.org/10.5281/zenodo.16358327. All other data that support the findings of this study are available from the corresponding author upon request.

\section*{Acknowledgments}
It is a pleasure to thank Moty Heiblum and Mike Zaletel for illuminating discussions. \textbf{Funding:} R.K. acknowledges support from the Dean of the Faculty and the Clore Foundation. Y.R. acknowledges the support from the Quantum Science and Technology Program 2021, the Schwartz Reisman Collaborative Science Program, the Gerald Schwartz and Heather Reisman Foundation, a research grant from the Goldfield Family Charitable Trust, the Minerva Foundation with funding from the Federal German Ministry for Education and Research, a research grant from the Estate of Hermine Miller, the Sheba Foundation, and Dweck Philanthropies, Inc., and the funding by the European Union (ERC, Anyons, 101163917). D.F.M. acknowledges support from the Israel Science Foundation (ISF) under grant 2572/21 and from the Minerva Foundation with funding from the Federal German Ministry for Education and Research. 

\section*{Author contributions}
R.K. fabricated the device. R.K., J.K., K.K., V.B, and A.I. helped in improving the device quality. K.W. and T.T. grew the hBN crystals. A.H. developed the measurement codes. R.K. and A.H. performed the measurements. R.K., A.H., M.Y., D.F.M., and Y.R. analyzed the measured data. M.Y. and D.F.M. developed the theoretical aspect. R.K., A.H., M.Y., D.F.M., and Y.R. authored the paper with input from all coauthors. Y.R. supervised the overall work done on the project.

\section*{Competing interests}
The authors declare no competing interests.
\newpage
\includepdf[pages={1-30}]{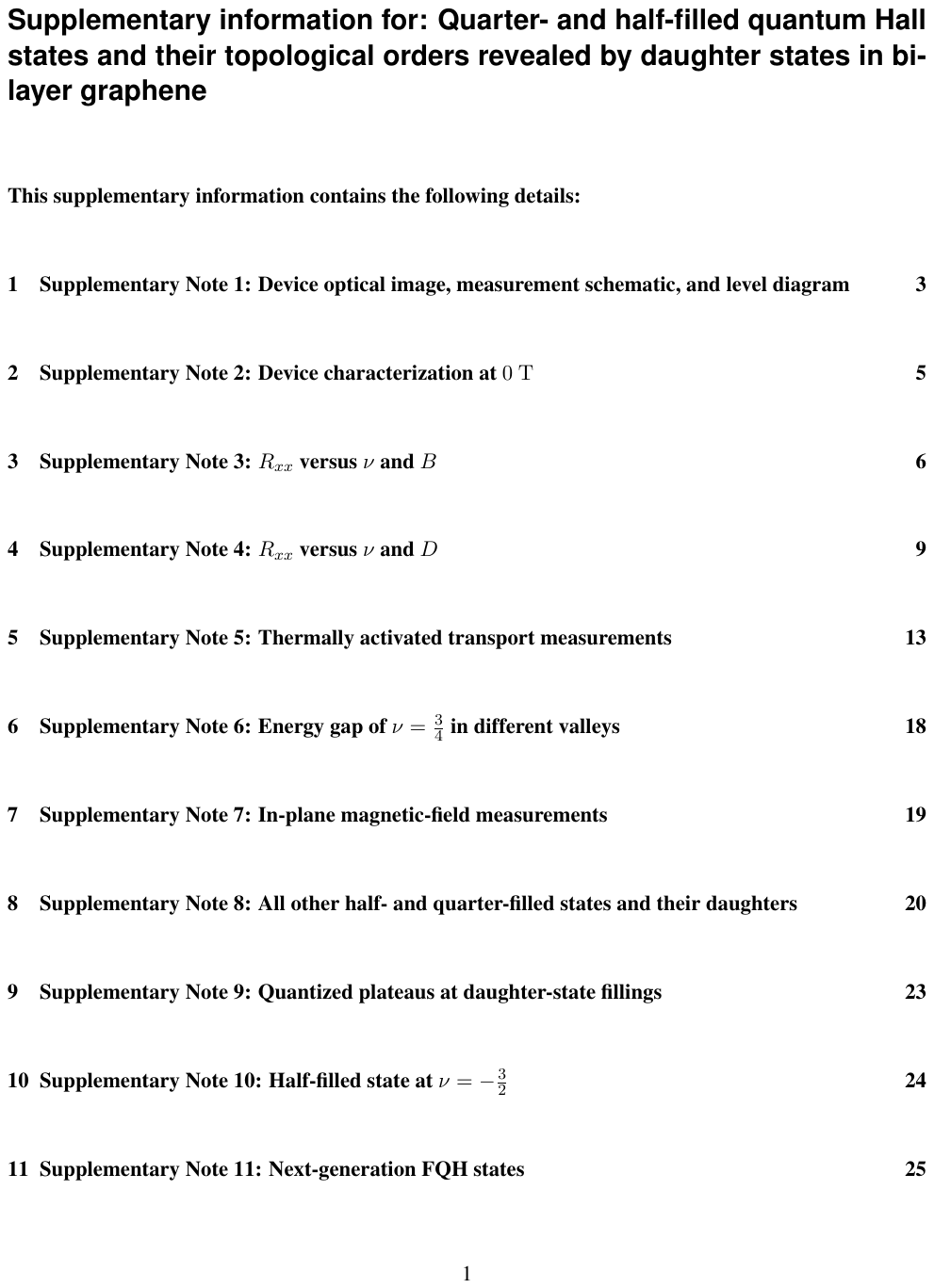}

\end{document}